\documentclass{pj}
\usepackage{graphicx}
\usepackage{amsmath,amssymb}

\begin{document}
\setcounter{page}{1}
\pjheader{July 2, 2020}

\title[iESC: iterative Equivalent Surface Current Approximation]
{iESC: iterative Equivalent Surface Current Approximation}  \footnote{\hskip-0.12in*\, Corresponding
author:~Shaolin~Liao~ (sliao5@iit.edu).} 
\footnote{\hskip-0.12in\textsuperscript{1} S. Liao is with Department of Electrical and Computer Engineering, Illinois Institute of Technology, Chicago, IL 60616 USA. \textsuperscript{2} L. Ou (oulu9676@gmail.com) is with College of Computer Science and Electronic Engineering, Hunan University, Changsha, Hunan, China  410082.}

\author{Shaolin~Liao\textsuperscript{*,1} and Lu~Ou\textsuperscript{2}}

\runningauthor{Liao and Ou}


\begin{abstract}
A novel iterative Equivalent Surface Current (iESC) algorithm has been developed to simulate the electromagnetic scattering of electrically large dielectric objects with relatively smooth surfaces. The iESC algorithm corrects the surface currents to compensate for the electromagnetic field deviation across the dielectric surface. Numerically validation has been performed with a dielectric sphere to show the performance of the iESC algorithm. The experimental result shows that it takes only a few iterations for the algorithm to increase the surface current accuracy by more than three orders of magnitude.
\end{abstract}


\setlength {\abovedisplayskip} {6pt plus 3.0pt minus 4.0pt}
\setlength {\belowdisplayskip} {6pt plus 3.0pt minus 4.0pt}

\

\section{Introduction}
\label{sec:intro}   
Electromagnetic scattering of dielectric objects has been considered as a challenging problem due to complicate dielectric surface geometries, various roughness and dielectric contrasts. Usually, rigorous Computational Electromagnetics (CEM) methods such as the Method of Moments (MoM) have to be used to obtain accurate solutions \cite{RWG}, \cite{Liao_Ping_Pong_APMC_2020}.  However, for electrically large objects in high-frequency electromagnetic frequencies from microwave \cite{liao_spectral-domain_2019}-\cite{Liao_ZIM_ATS_2019} and millimeter wave \cite{Liao_Thesis_Proquest_2008}-\cite{Liao_overmode_JEMWA_2008} to optics \cite{Liao_ZIM_arXiv_2020}-\cite{Liao_X_Ray_arXiv_2020}, efficient CEM algorithms are required. In this paper, a novel iterative Equivalent Surface Current (iESC) approximation algorithm is developed to for electrically large dielectric objects of relatively smooth surfaces.

\begin{figure}[th]
 \centering
 \includegraphics[width=0.35\textwidth]{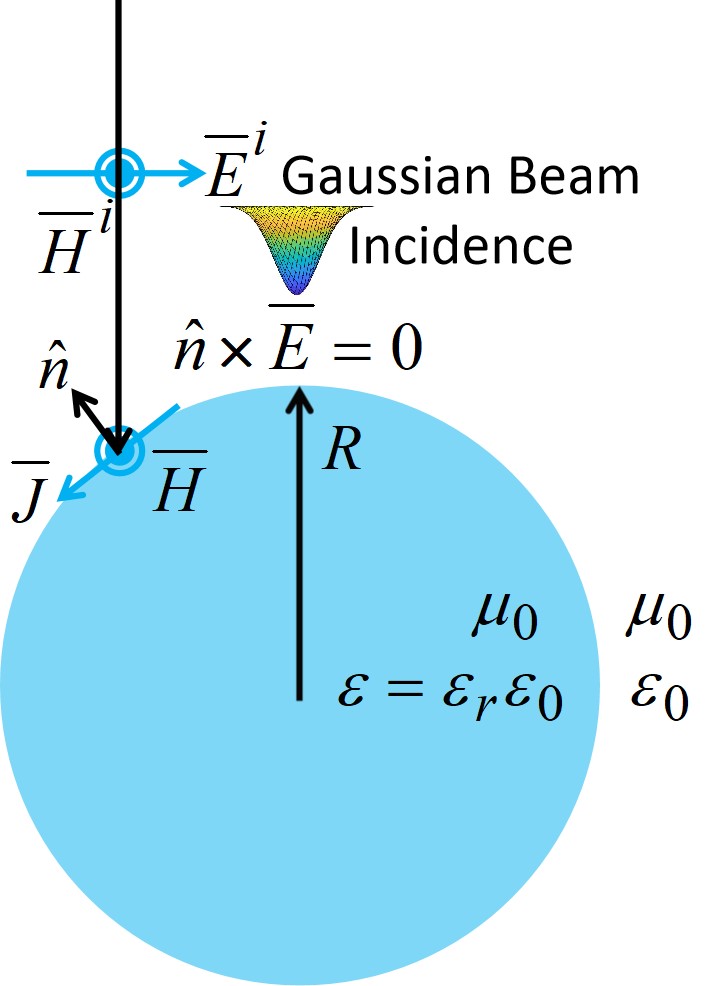}
\caption{The iESC approximation of a dielectric sphere with a permittivity of $\epsilon = \epsilon_r \epsilon_0 = 2 \epsilon_0$.}
\label{fig:problem}
\end{figure}

\section{Problem Formulation}\label{sec:problem}
The electromagnetic scattering problem from a dielectric surface is shown in Fig. \ref{fig:problem}, which can be formulated with the surface currents \cite{Liao_Thesis_Proquest_2008}. In general, the unknown surface currents can only be solved through rigorous CEM methods such as MoM by imposing the continuities of the electromagnetic field across the dielectric surface as follows,
\begin{flalign}\label{eqn:BC}
&\hat{n} \times  \delta \overline{E} (x, y)  = \hat{n} \times \left\{\overline{E}^i (x, y) +  \overline{E}^+ (x, y) -  \overline{E}^- (x, y)  \right\}  =0, \nonumber \\
& \hat{n} \times  \delta \overline{H} (x, y) = \hat{n} \times \left\{\overline{H}^i (x, y) +  \overline{H}^+ (x, y) -  \overline{H}^- (x, y)  \right\}  =0, 
\end{flalign}
where $\hat{n}$ is the unit surface normal of the dielectric surface; also, the superscripts $\pm$ and $i$ denote the scattering electric/magnetic field on the upper/lower surface and the incident electric field, respectively.

\begin{figure}[th]
 \centering
 \includegraphics[width=0.55\textwidth]{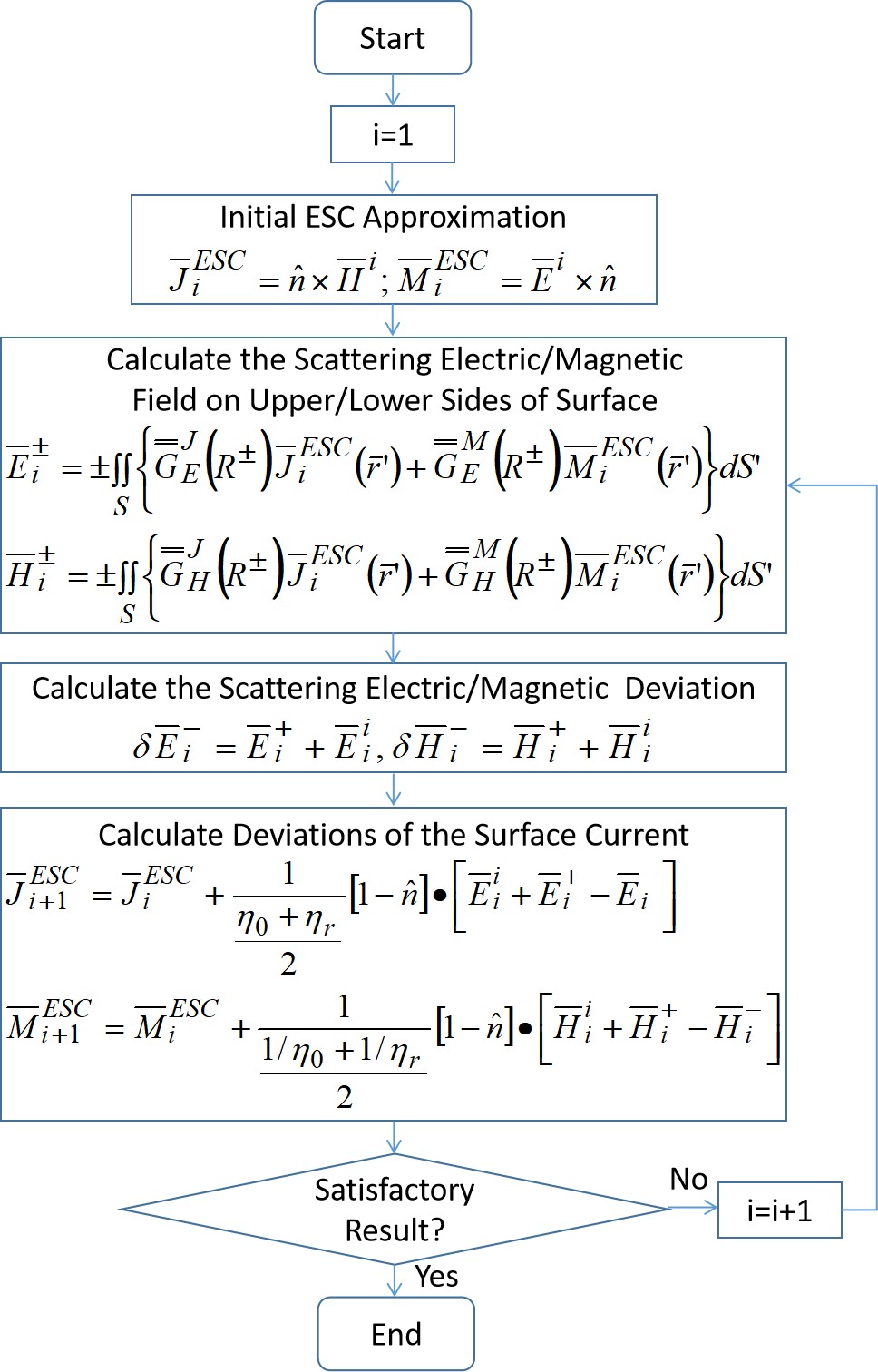}
\caption{Algorithm of the iESC approximation.}
\label{fig:algorithm}
\end{figure}

\section{The Scattering Electromagnetic Field}
 The scattering electric/magnetic field $\overline{E}^\pm/\overline{H}^\pm$ on the upper/lower dielectric surface can be expressed in terms of convolution of the surface currents $\overline{J}/\overline{M}$ and the electric dyadic Green's functions as follows \cite{Liao_Thesis_Proquest_2008},
\begin{flalign}\label{eqn:field_convolution}
\overline{E}^\pm = \overline{\overline{G}}_E^{J\pm} \circledast \overline{J}(\overline{r}') + \overline{\overline{G}}_E^{M\pm} \circledast \overline{M}(\overline{r}')   =   \iint\limits_{S} \left\{ \overline{\overline{G}}_E^J \left(R^\pm\right) \overline{J}(\overline{r}') +  \overline{\overline{G}}_E^M \left(R^\pm\right) \overline{M}(\overline{r}') \right\} dS',  \\
\overline{H}^\pm  = \overline{\overline{G}}_H^{J\pm} \circledast \overline{J}(\overline{r}') + \overline{\overline{G}}_H^{M\pm} \circledast \overline{M}(\overline{r}')  =   \iint\limits_{S} \left\{ \overline{\overline{G}}_H^J \left(R^\pm\right) \overline{J}(\overline{r}') +  \overline{\overline{G}}_H^M \left(R^\pm\right) \overline{M}(\overline{r}') \right\} dS', \nonumber
\end{flalign}
where $\circledast$ denotes the convolution; $R^\pm = r^\pm- r'; r^\pm=|\overline{r}^\pm|, r'=|\overline{r}'|$ with $\overline{r}^\pm$ and $\overline{r}'$ being the observation points on the upper/lower surface and the source point respectively; $\omega$ is the angular frequency; $\mu$ and $\epsilon$ are the permeability and permittivity.
 
\section{The iESC Approximation}
Starting with the initial guess of the surface currents of $\overline{J}_1^{ESC}/\overline{M}_1^{ESC}$, the deviation of the electromagnetic field of Eq. (\ref{eqn:BC}) is given by,
\begin{flalign}\label{eqn:dE}
&   \delta \overline{E}_1    = \overline{E}^i   +  \left[\overline{\overline{G}}_E^{J+} \circledast \overline{J}_1^{ESC}  + \overline{\overline{G}}_E^{M+} \circledast \overline{M}_1^{ESC}  \right]  - \left[  \overline{\overline{G}}_E^{J-} \circledast \overline{J}_1^{ESC} + \overline{\overline{G}}_E^{M-} \circledast \overline{M}_1^{ESC}  \right] \nonumber \\ 
& \delta \overline{H}_1    = \overline{H}^i   +  \left[ \overline{\overline{G}}_H^{J+} \circledast \overline{J}_1^{ESC} + \overline{\overline{G}}_H^{M+} \circledast \overline{M}_1^{ESC}  \right] -   \left[  \overline{\overline{G}}_H^{J-} \circledast \overline{J}_1^{ESC}  + \overline{\overline{G}}_H^{M-} \circledast \overline{M}_1^{ESC}  \right], 
\end{flalign}

Approximating the local electromagnetic field as local plane wave, the electric field deviation $ \delta \overline{E}_1$ is related to the magnetic field deviation $ \delta \overline{H}_1$ as follows,
\begin{flalign}\label{eqn:dH0}
 \delta \overline{E}_1     =   \eta^\pm  \delta \overline{H}_1   \times \hat{n}; \ \  \delta \overline{H}_1    =    \hat{n}   \times  \frac{\delta \overline{E}_1}{\eta^\pm}; \ \ \eta^\pm = \sqrt{\frac{\mu}{\epsilon^\pm}}.
\end{flalign}
from which the following average electric field deviation and magnetic field deviation can be defined,
\begin{flalign}\label{eqn:dH}
 & \delta \overline{E}_1^{ESC} = \frac{1}{2} \left(  \delta \overline{E}_1^+ +  \delta \overline{E}_1^-\right) = \frac{1}{2}\left( \eta^+ + \eta^-  \right) \delta \overline{H}_1  \times \hat{n} , \nonumber \\
 & \delta \overline{H}_1^{ESC}  =  \frac{1}{2} \left(  \delta \overline{H}_1^+ +  \delta \overline{H}_1^-\right) =  \frac{1}{2}\left( \frac{1}{\eta^+} + \frac{1}{\eta^-}  \right) \hat{n}   \times  \delta \overline{E}_1,
\end{flalign}

To compensate the deviations of the magnetic field deviation $\delta  \overline{H}_1^{ESC}$ and the electric field deviation $\delta \overline{E}_1^{ESC}$, the surface current deviations $\delta \overline{J}_1^{ESC}$ and $\delta \overline{M}_1^{ESC}$ can be corrected as follows \cite{Liao_IPO_APMC_2020}, 
\begin{flalign}\label{dJdM}
\delta \overline{J}_1^{ESC} &= \frac{1}{\frac{1}{2}\left( \eta^+ + \eta^-  \right)} \left[ 1- \hat{n} \right] \bullet \delta \overline{E}_1   = \frac{1}{\frac{1}{2}\left( \eta^+ + \eta^-  \right)} \left[ 1- \hat{n} \right] \bullet \left\{ \overline{E}^i   +  \overline{E}_1^+   -  \overline{E}_1^-    \right\}, \nonumber \\
\delta \overline{M}_1^{ESC} &= \frac{1}{\frac{1}{2}\left( \frac{1}{\eta^+} + \frac{1}{\eta^-}  \right)} \left[ 1- \hat{n} \right] \bullet \delta \overline{H}_1   = \frac{1}{\frac{1}{2}\left( \frac{1}{\eta^+} + \frac{1}{\eta^-}   \right)} \left[ 1- \hat{n} \right] \bullet \left\{ \overline{H}^i   +  \overline{H}_1^+   -  \overline{H}_1^-    \right\}.
\end{flalign}  
  
The corrections of the surface currents in Eq. (\ref{dJdM}) repeat until the solutions converge to their minimum values.

\section{Algorithm}
Fig. \ref{fig:algorithm} shows the algorithm of the improved iESC approximation: it starts with the following initial guess;
\begin{flalign}
\overline{J}_1^{ESC}  = \hat{n} \times \overline{H}^i, \ \ \overline{J}_1^{ESC}  =  \overline{E}^i \times \hat{n}.
\end{flalign}

Then it calculates the scattering electromagnetic field according to Eq. (\ref{eqn:field_convolution}).

After that, corrections of the surface currents in Eq. (\ref{dJdM}) repeat until the minimum deviations are obtained.

\begin{figure}[t]
 \centering
 \includegraphics[width=0.95\textwidth]{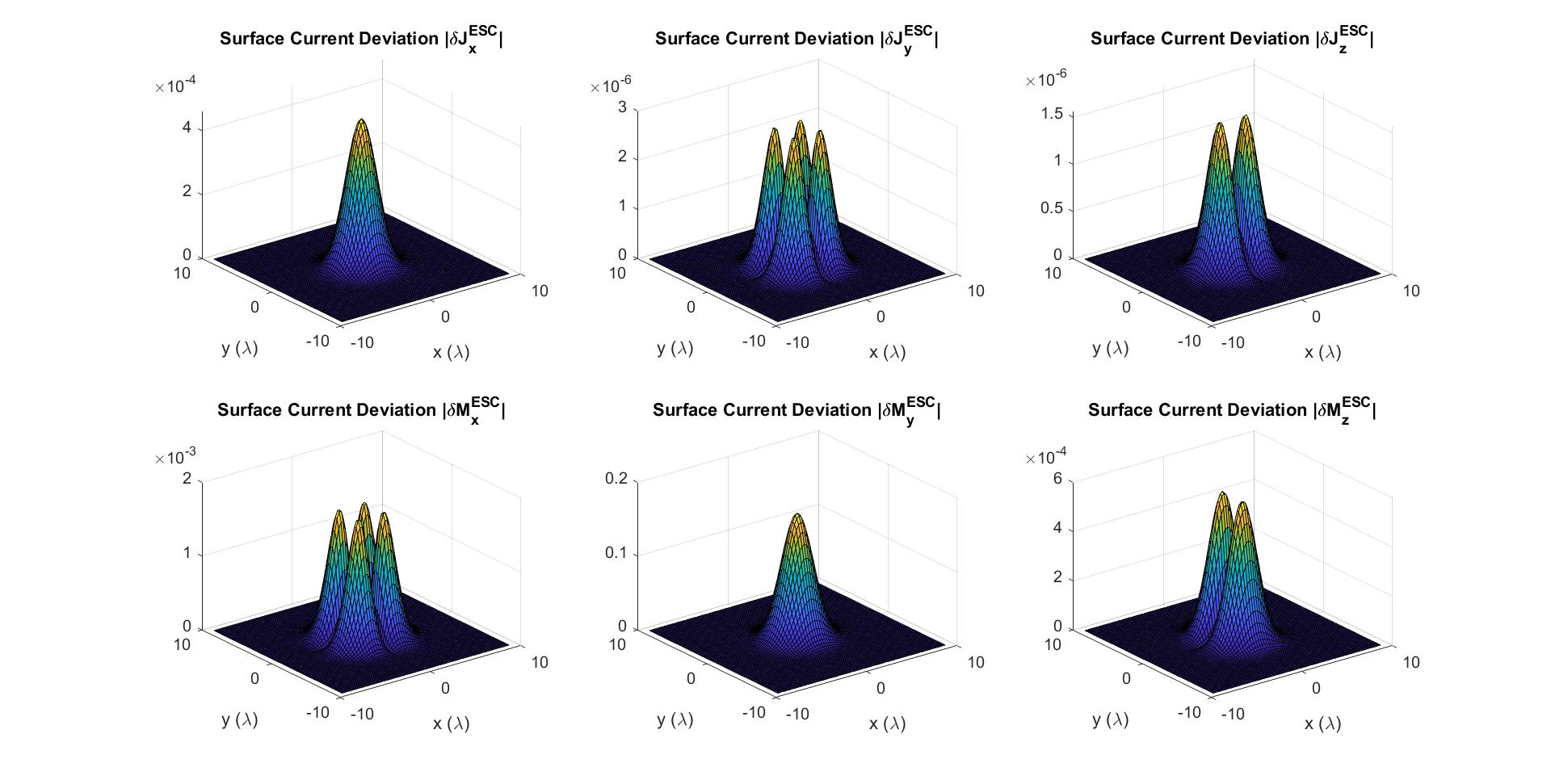}
\caption{First-iteration iESC surface currents deviations for a dielectric sphere with a  radius of $R = 250 \lambda$: top row) $\left| \delta \overline{J}_1^{ESC} \right|$, from left to right are $\hat{x}$, $\hat{y}$ and $\hat{z}$ components respectively; and top row) $\left| \delta \overline{M}_1^{ESC} \right|$, from left to right are $\hat{x}$, $\hat{y}$ and $\hat{z}$ components respectively.}
\label{fig:JM1}
\end{figure}

\begin{figure}[th]
 \centering
 \includegraphics[width=0.95\textwidth]{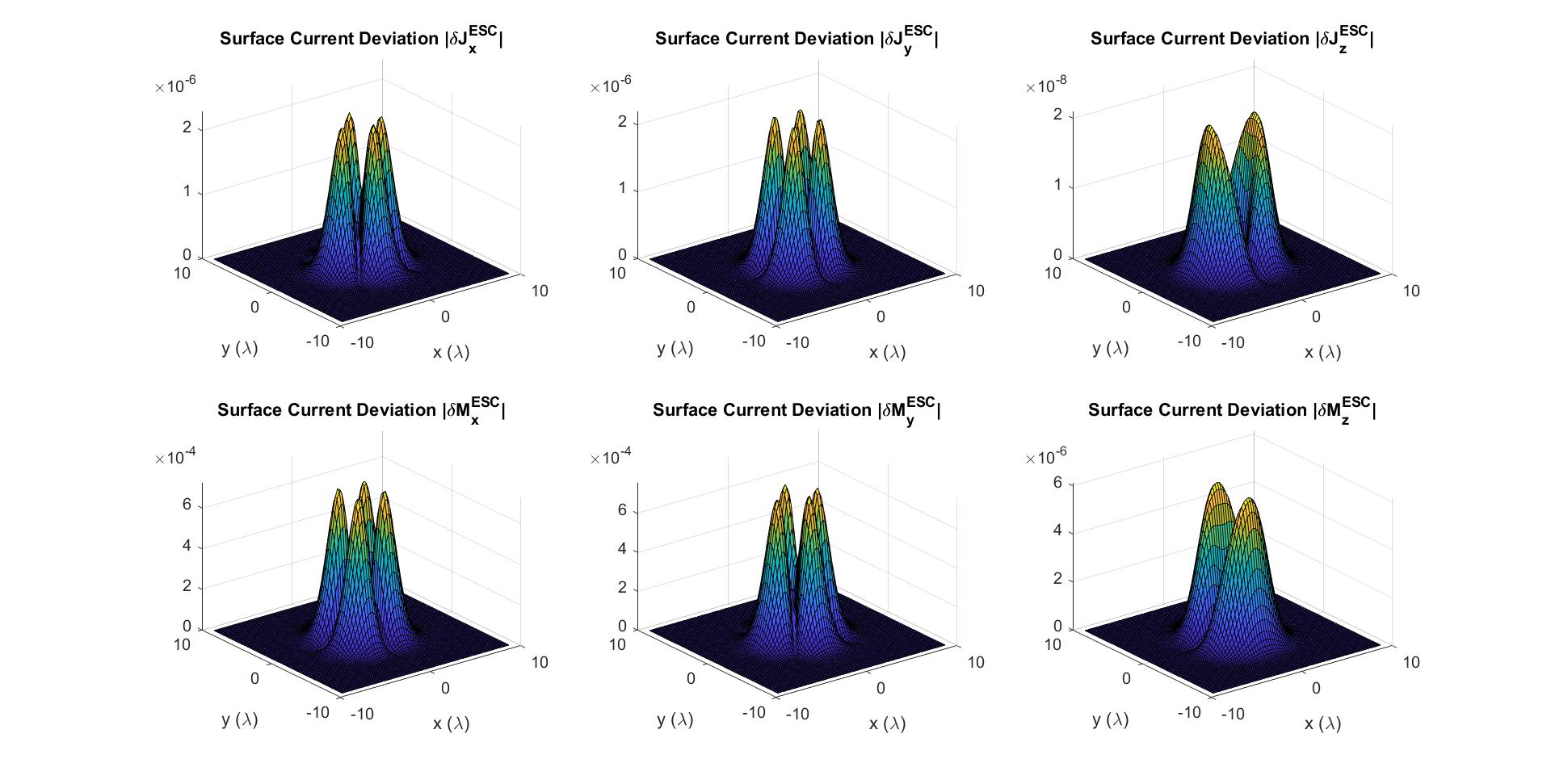}
\caption{Second-iteration iESC surface currents deviations for a dielectric sphere with a  radius of $R = 250 \lambda$: top row) $\left| \delta \overline{J}_1^{ESC} \right|$, from left to right are $\hat{x}$, $\hat{y}$ and $\hat{z}$ components respectively; and top row) $\left| \delta \overline{M}_1^{ESC} \right|$, from left to right are $\hat{x}$, $\hat{y}$ and $\hat{z}$ components respectively.}
\label{fig:JM2}
\end{figure}

\begin{figure}[th]
 \centering
 \includegraphics[width=0.95\textwidth]{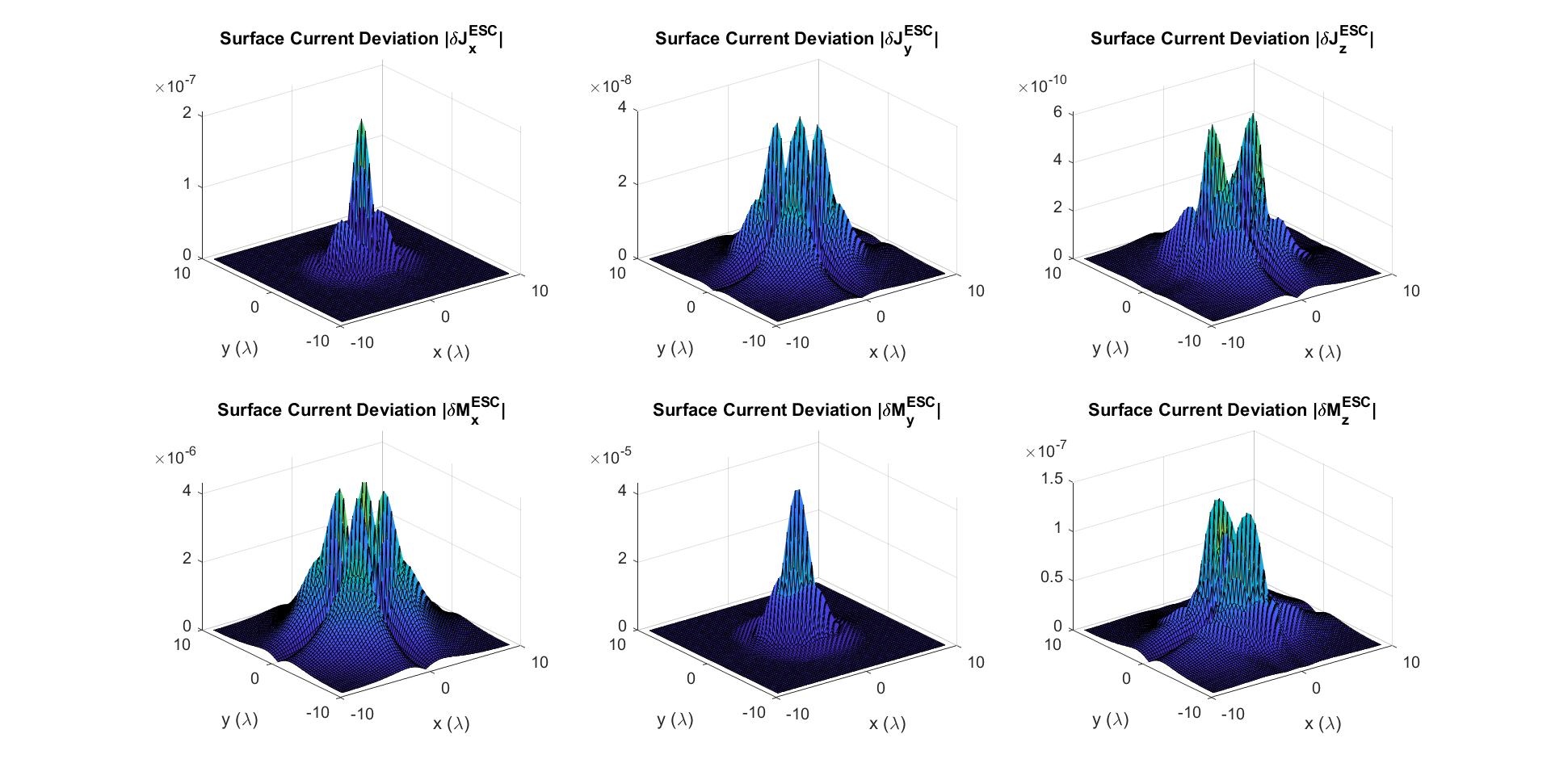}
\caption{Third-iteration iESC surface currents deviations for a dielectric sphere with a  radius of $R = 250 \lambda$: top row) $\left| \delta \overline{J}_1^{ESC} \right|$, from left to right are $\hat{x}$, $\hat{y}$ and $\hat{z}$ components respectively; and top row) $\left| \delta \overline{M}_1^{ESC} \right|$, from left to right are $\hat{x}$, $\hat{y}$ and $\hat{z}$ components respectively.}
\label{fig:JM3}
\end{figure}

\begin{figure}[th]
 \centering
 \includegraphics[width=1.\textwidth]{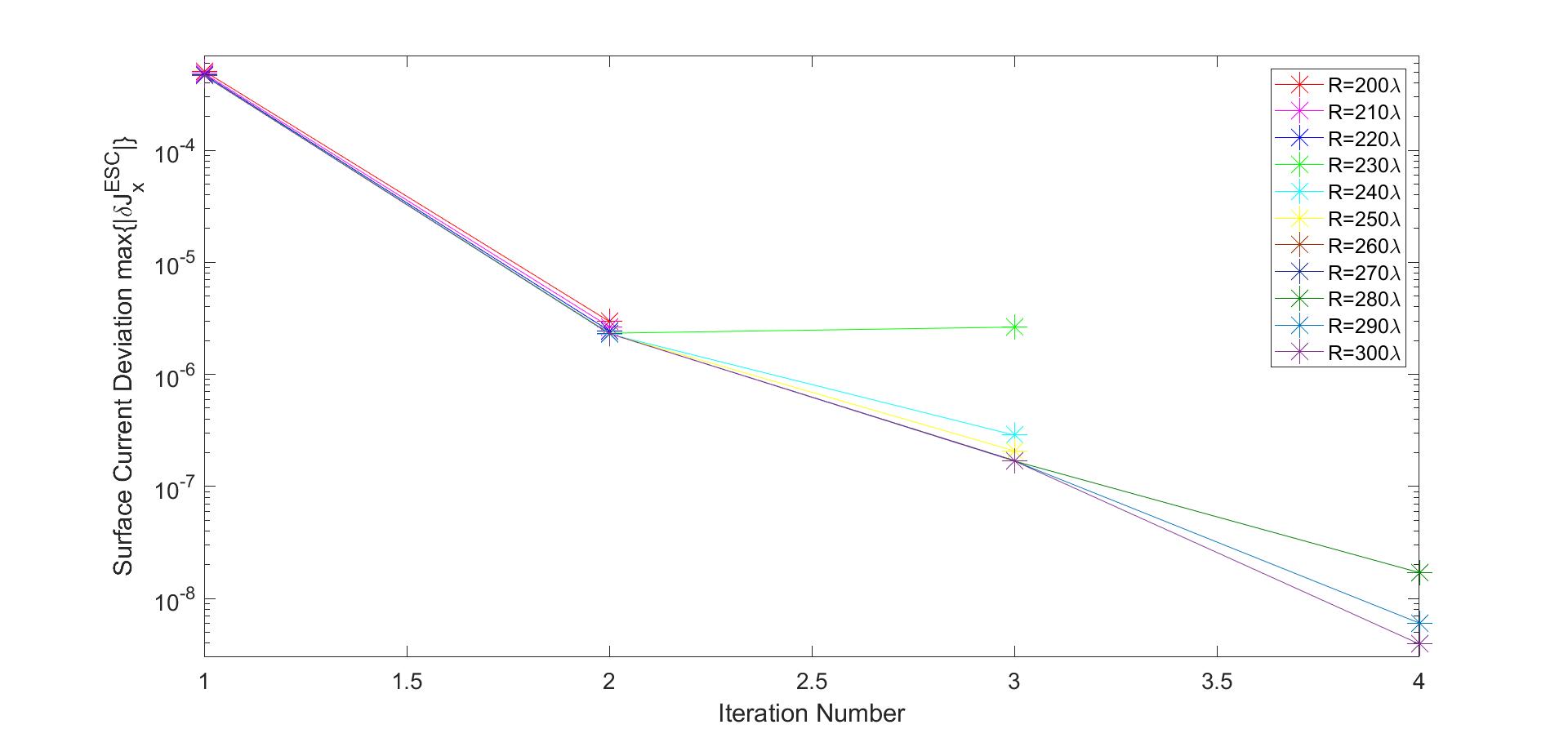}
\caption{iESC surface current deviation $max\left\{\left|d\overline{J}_x^{ESC} \right|\right\}$ for a dielectric sphere of various  radii of $R = [200, 300]\lambda$ at different iterations.}
\label{fig:dJ}
\end{figure}

\begin{figure}[th]
 \centering
 \includegraphics[width=1.\textwidth]{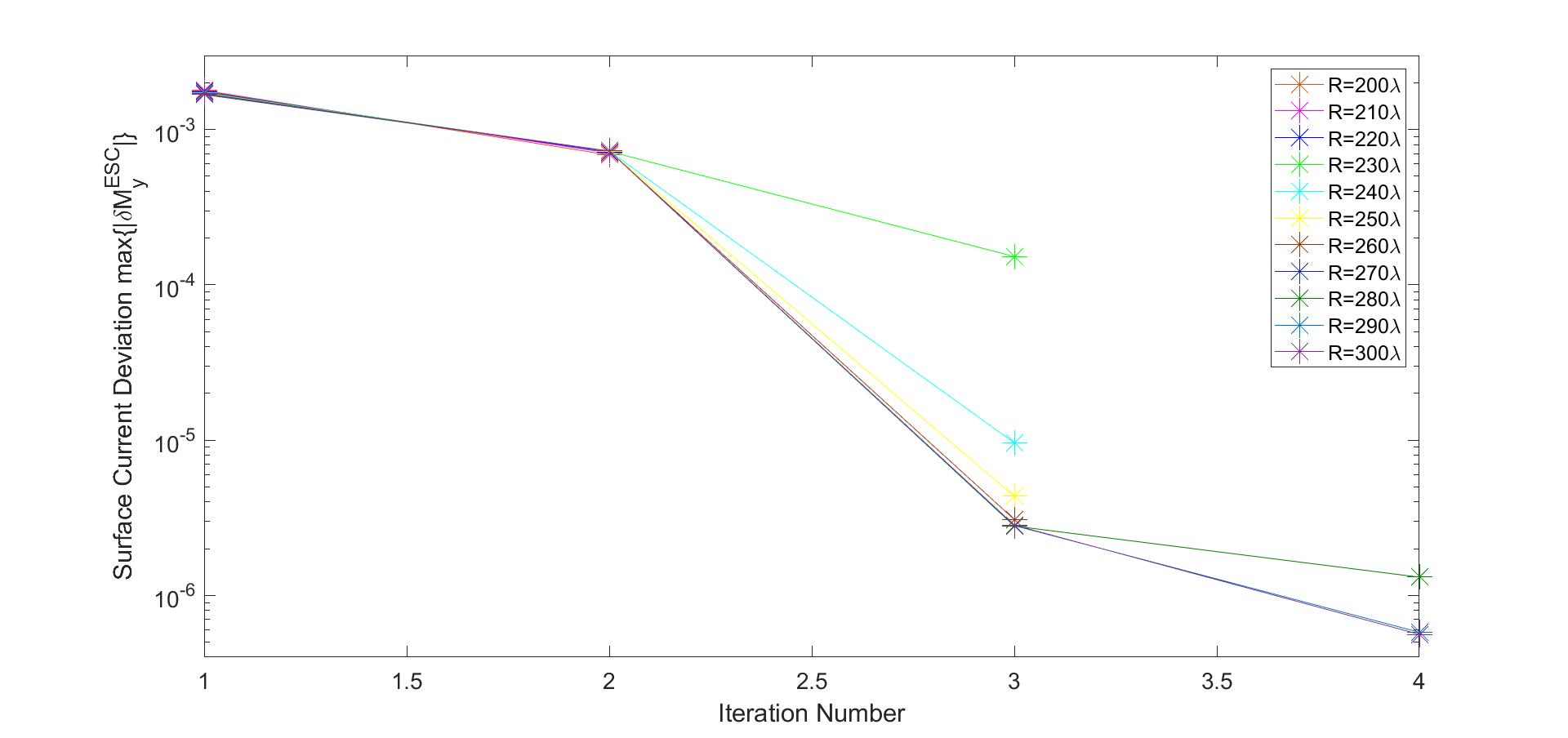}
\caption{iESC surface current deviation $max\left\{\left|d\overline{M}_y^{ESC} \right|\right\}$ for a dielectric sphere of various radii of $R = [200, 300]\lambda$ at different iterations.}
\label{fig:dM}
\end{figure}

\section{Numerical Validation}
To show the efficiency of the iESC approximation, numerical validation has been performed with a dielectric sphere ($\epsilon_r = 2$) of various radii $R = [200, 300]\lambda$ and with a 
Gaussian beam of waist $w = 2 \lambda$ as the incidence wave.

Fig. \ref{fig:JM1}, Fig. \ref{fig:JM2} and Fig. \ref{fig:JM3} show the deviations of the surface currents $|\overline{J}_i^{ESC}|$ and   $|\overline{M}_i^{ESC}|$ for the first, second and third iterations respectively, from which it can be seen that the iESC approximation only takes 3 iterations to converge.

Also, Fig. \ref{fig:dJ} and Fig. \ref{fig:dM} show the convergence of the iESC approximation, from which it can be seen that the accuracy of the ESC has been increased by more than three orders of magnitude.

\section{Conclusion}\label{sec:con}
We have developed an efficient iESC algorithm for simulation of electromagnetic scattering of electrically large dielectric objects whose surface are relatively smooth. Through iterative correction of the surface currents, the iESC algorithm can increase the solution accuracy by more than three orders of magnitude within a few iterations. The iESC algorithm has many potential important applications such as electromagnetic scattering of ocean surfaces, ice layers, and soils.

\ack
The work is partially supported by China Postdoctoral Science Foundation No. 2020M672488.

\end{document}